# Simple proof of gauge invariance for the S-matrix element of strong-field photoionization

Jarosław H Bauer

Katedra Fizyki Teoretycznej Uniwersytetu Łódzkiego,

Ul. Pomorska 149/153, 90-236 Łódź, Poland

E-mail:bauer@uni.lodz.pl

**Abstract.** The relationship between the length gauge (LG) and the velocity gauge (VG) exact forms of the photoionization probability amplitude is considered. Our motivation for this paper comes from applications of the Keldysh-Faisal-Reiss (KFR) theory, which describes atoms (or ions) in a strong laser field (in the nonrelativistic approach, in the dipole approximation). On the faith of a certain widely-accepted assumption, we present a simple proof that the well-known LG form of the exact photoionization (or photodetachment) probability amplitude is indeed the gauge-invariant result. In contrast, to obtain the VG form of this probability amplitude, one has to either (i) neglect the well-known Göppert-Mayer exponential factor (which assures gauge invariance) during all the time evolution of the ionized electron or (ii) put some conditions on the vector potential of the laser field.

The Keldysh-Faisal-Reiss (KFR) theory [1-4] describes nonresonant multiphoton processes such as above threshold detachment of ions and above threshold ionization of atoms in an intense electromagnetic (laser) field. This theory utilizes the $S$-matrix theory, which is in principle exact, but in practice one has to use approximate wavefunctions to evaluate the $S$-matrix elements for bound-free transitions. This is due to non existence of the general analytical solution to the time-dependent Schrödinger equation (TDSE) for a charged particle interacting with both an attractive potential (for example the Coulomb one) and an electromagnetic plane-wave field even in the simplest case. In this paper we consider the outgoing nonrelativistic electron in the dipole (or long wavelength) approximation. The electron is initially bound by any attractive potential. Similar assumptions about the electronic motion were made in the pioneering papers [1-3,5], which utilize the Gordon-Volkov final state wavefunctions [6,7]. These wavefunctions, which completely omit the attractive potential, are good enough only for sufficiently strong laser fields (particularly of circular polarization) [3,4]. Such simple description is not sufficient for each present-day strong-field ionization or detachment experiment, including the numerical ones. However, any generalization of the KFR theory, accounting for magnetic-field component of the laser field [8] or relativistic effects, should reduce to the correct nonrelativistic dipole approximation limit for lower radiation intensities. It was very well known long ago that the length gauge (LG) (used by Keldysh) and the velocity gauge (VG) (used by Reiss) theories give different results (see, for example, [4,9-16]). Recently this problem attracted quite considerable interest as well [17-23], where some advantages of the LG theory over the VG one were mostly stressed. According to Frolov *et al* [24], the $S$-matrix approach does not take into account the level shifting (dynamic Stark effect) and the laser-induced level width of an initial state.

Let us assume that the laser field (of any elliptical polarization), which is adiabatically turned on and off at asymptotic times ($t \to \pm\infty$) changes arbitrarily in time (in the KFR theory a monochromatic laser field of constant amplitude is considered, but in this paper this condition is superfluous). The laser field can be described by the vector potential $\vec{A}$ (in the Coulomb gauge), for which one usually assumes that:

$$\lim_{t \to \pm\infty} \vec{A}(t) = \vec{0} \ . \tag{1}$$

The lack of spatial dependence in the vector potential $\vec{A}(t)$ means that the dipole approximation (which is roughly valid, if the wavelength of the incident radiation is much larger than the atomic size, but see [8] for more details) has been applied. The magnetic-field component of the laser field is zero and the electric-field component is given by

$$\vec{F}(t) = -\frac{1}{c}\frac{\partial \vec{A}}{\partial t} \ . \tag{2}$$

At this place let us only note (see equation (2)) that the condition (1) is stronger than the condition: $\lim_{t \to \pm\infty} \vec{F}(t) = \vec{0}$. The one-electron atom or ion in the laser field is described by the following Hamiltonian

$$H = H_A + H_I \equiv \frac{\vec{p}^2}{2m} + V(\vec{r}) + H_I , \qquad (3)$$

where $H_A$ is the atomic Hamiltonian, and the interaction Hamiltonian is given either in the $\vec{p}\vec{A}$ form:

$$H_I^{pA} = \frac{-e}{mc}\vec{A}(t)\vec{p} + \frac{e^2}{2mc^2}\vec{A}(t)^2 , \qquad (4)$$

($e < 0$ is the charge of an electron and $\vec{p} = -i\hbar\vec{\nabla}$) or in the $\vec{d}\vec{E}$ form:

$$H_I^{dE} = -e\vec{r}\vec{F}(t) . \qquad (5)$$

Using the usual rules of quantum mechanics for both wavefunctions and operators ($\Psi' = \hat{U}\Psi$, $\hat{O}' = \hat{U}\hat{O}\hat{U}^{-1}$), the TDSE (for the atom in the laser field) can be transformed from either form to the other one by a certain unitary transformation. It is very well-known (see for instance [25-27]) that within the dipole approximation of the laser field the above mentioned two descriptions are equivalent. Let $\Psi(\vec{r},t)$ be the exact solution of the TDSE with the Hamiltonian (3) in either form, and $\Phi(\vec{r},t)$ be the exact solution of the TDSE with the Hamiltonian $H_A$. If the subscripts $i, f$ denote initial and final states, one can define two equivalent forms of the exact $S$-matrix:

$$S_{fi} = \lim_{t \to +\infty} \left(\Phi_f, \Psi_i^{(+)}\right), \qquad \text{where} \qquad \lim_{t \to -\infty} \Psi_i^{(+)} = \Phi_i , \qquad (6)$$

$$S_{fi} = \lim_{t \to -\infty} \left(\Psi_f^{(-)}, \Phi_i\right), \qquad \text{where} \qquad \lim_{t \to +\infty} \Psi_f^{(-)} = \Phi_f , \qquad (7)$$

which are called the direct time form and the reversed time form respectively (the round brackets denote the overlap of two wavefunctions). In both papers [1,3] the latter one was utilized, so we will set about doing it. In this case including the boundary condition for $t \to +\infty$ (the second of equations (7), see also Section 2 of [28]), one obtains

$$(S-1)_{fi} = S_{fi} - \delta_{fi} = \lim_{t \to -\infty} \left( \Psi_f^{(-)}, \Phi_i \right) - \lim_{t \to +\infty} \left( \Psi_f^{(-)}, \Phi_i \right) = \int_{\infty}^{-\infty} dt \frac{\partial}{\partial t} \left( \Psi_f^{(-)}, \Phi_i \right)$$

$$= -\int_{-\infty}^{\infty} dt \left( \frac{\partial}{\partial t} \Psi_f^{(-)}, \Phi_i \right) - \int_{-\infty}^{\infty} dt \left( \Psi_f^{(-)}, \frac{\partial}{\partial t} \Phi_i \right) = -\int_{-\infty}^{\infty} dt \left( \frac{1}{i\hbar} (H_A + H_I) \Psi_f^{(-)}, \Phi_i \right)$$

$$- \int_{-\infty}^{\infty} dt \left( \Psi_f^{(-)}, \frac{1}{i\hbar} H_A \Phi_i \right) = -\frac{i}{\hbar} \int_{-\infty}^{\infty} dt \left( \Psi_f^{(-)}, H_I \Phi_i \right)$$

, (8)

where the hermiticity of $H_A$ and $H_I$ has been used. The crucial point of this paper is the following observation: only in the LG equations (6), (7) are true probability amplitudes of ionization (or detachment), if the wavefunctions $\Phi_f$ and $\Phi_i$ denote the "textbook" wavefunctions without any additional phase factors. There are two reasons which make us to think so. The first one is the following. The asymptotic reference states should be eigenstates of the gauge-invariant energy operator in the absence of the laser field. In appendix we show that this implies that only in the LG this asymptotic reference states have the form of the "textbook" wavefunctions $\Phi_f$ or $\Phi_i$. To get the probability amplitude of ionization in the VG, first one has to transform these wavefunctions to this gauge according to

$$\Psi^{pA}(\vec{r},t) = \hat{U}(\vec{r},t) \Psi^{dE}(\vec{r},t) , \qquad \text{where} \qquad \hat{U}(\vec{r},t) = \exp\left( \frac{ie}{\hbar c} \vec{r} \vec{A}(t) \right) \qquad (9)$$

is the Göppert-Mayer unitary operator [29], which has just been mentioned above and which assures gauge invariance. This exponential operator transforms the TDSE with the Hamiltonian (3) (both wavefunctions and operators) from the LG to the VG [25-27]. Necessity of multiplying the wavefunction $\Psi^{pA}(\vec{r},t)$ by the operator $\hat{U}^+(\vec{r},t) = \exp\left( -ie/\hbar c) \vec{r} \vec{A}(t) \right)$ before projecting on an eigenstate of the atomic Hamiltonian was already noticed by many authors (which is the second reason of the two mentioned above) long ago. It took place in various contexts [30-42] (of course this list of references is not complete). In this way one obtains the overlap, which is gauge-invariant and can be treated as an instantaneous probability amplitude of ionization (or detachment). This is equivalent to multiplying the wavefunction $\Psi^{dE}(\vec{r},t)$ by the operator $\hat{U}(\vec{r},t)$, when working in the VG. For sufficiently weak electric fields described by $\vec{A}(t)$ one can sometimes approach $\hat{U}(\vec{r},t)$ by unity, but for strong fields putting $\hat{U}(\vec{r},t) \cong 1$ is certainly not justified. For example, for the $H(1s)$ atom, if $|\vec{r}| \cong 1$ *a.u.* the exponent in equation (9) is of the order of $i\sqrt{z_1}$, where $z_1$ is the Reiss (dimensionless) intensity parameter [3]. At the lower applicability limit of the KFR theory $z_1 = 10$ and it grows with increasing intensity. The counterparts of equations (6), (7) in the VG are the following:

$$S_{fi} = \lim_{t \to +\infty} \left( \hat{U} \Phi_f, \Psi_i^{(+)pA} \right) = \lim_{t \to +\infty} \left( \Phi_f, \hat{U}^+ \Psi_i^{(+)pA} \right), \quad \text{where} \quad \lim_{t \to -\infty} \Psi_i^{(+)pA} = \hat{U} \Phi_i, \quad (10)$$

$$S_{fi} = \lim_{t \to -\infty} \left( \Psi_f^{(-)pA}, \hat{U} \Phi_i \right) = \lim_{t \to -\infty} \left( \hat{U}^+ \Psi_f^{(-)pA}, \Phi_i \right), \quad \text{where} \quad \lim_{t \to +\infty} \Psi_f^{(-)pA} = \hat{U} \Phi_f. \quad (11)$$

To obtain correct and exact result in the VG (for the reversed time $S$-matrix) we have to start from equations (11).

$$(S-1)_{fi} = S_{fi} - \delta_{fi} = \lim_{t \to -\infty} \left( \Psi_f^{(-)pA}, \hat{U} \Phi_i \right) - \lim_{t \to +\infty} \left( \Psi_f^{(-)pA}, \hat{U} \Phi_i \right) = \int_{\infty}^{-\infty} dt \frac{\partial}{\partial t} \left( \hat{U}^+ \Psi_f^{(-)pA}, \Phi_i \right)$$

$$= -\int_{-\infty}^{\infty} dt \left( \frac{\partial \hat{U}^+}{\partial t} \Psi_f^{(-)pA}, \Phi_i \right) - \int_{-\infty}^{\infty} dt \left( \hat{U}^+ \frac{\partial \Psi_f^{(-)pA}}{\partial t}, \Phi_i \right) - \int_{-\infty}^{\infty} dt \left( \hat{U}^+ \Psi_f^{(-)pA}, \frac{\partial \Phi_i}{\partial t} \right)$$

$$= -\int_{-\infty}^{\infty} dt \left( \frac{1}{i\hbar} H_I^{dE} \hat{U}^+ \Psi_f^{(-)pA}, \Phi_i \right) - \int_{-\infty}^{\infty} dt \left( \hat{U}^+ \frac{1}{i\hbar} (H_A + H_I^{pA}) \Psi_f^{(-)pA}, \Phi_i \right) - \int_{-\infty}^{\infty} dt \left( \hat{U}^+ \Psi_f^{(-)pA}, \frac{1}{i\hbar} H_A \Phi_i \right)$$

$$. (12)$$

Using equations (2), (5) and the hermiticity of $H_A$, we can add the last three integrals in equation (12) and obtain

$$(S-1)_{fi} = \frac{1}{i\hbar} \int_{-\infty}^{\infty} dt \left( \left( H_I^{dE} \hat{U}^+ + \left[ \hat{U}^+, H_A \right] + \hat{U}^+ H_I^{pA} \right) \Psi_f^{(-)pA}, \Phi_i \right). \quad (13)$$

The calculation of a commutator in the above equation is elementary and gives:

$$\left[ \hat{U}^+, H_A \right] = -\hat{U}^+ H_I^{pA}. \quad (14)$$

In this way, after substitution of equation (14) into equation (13), the Hamiltonian $H_I^{pA}$ disappears from the $S$-matrix element in spite of making calculations in the VG. Finally, because $H_I^{dE}$ is a hermitian operator, the exact probability amplitude of ionization takes the form

$$(S-1)_{fi} = \frac{1}{i\hbar} \int_{-\infty}^{\infty} dt \left( H_I^{dE} \hat{U}^+ \Psi_f^{(-)pA}, \Phi_i \right) = \frac{1}{i\hbar} \int_{-\infty}^{\infty} dt \left( \Psi_f^{(-)dE}, H_I^{dE} \Phi_i \right), \tag{15}$$

which is identical with the LG result derived in equation (8). In this way, for the exact probability amplitude of ionization, gauge invariance is preserved. By analogy, starting from equations (6) and (10) respectively, and making similar calculations, one can check, that the gauge-invariant direct time form of the $S$-matrix is the following

$$(S-1)_{fi} = \frac{1}{i\hbar} \int_{-\infty}^{\infty} dt \left( \Phi_f, H_I^{dE} \Psi_i^{(+)dE} \right). \tag{16}$$

In the light of the above calculations it is obvious that the well-known starting point of the VG KFR theory [3]

$$(S-1)_{fi} = \frac{1}{i\hbar} \int_{-\infty}^{\infty} dt \left( \Psi_f^{(-)pA}, H_I^{pA} \Phi_i \right), \tag{17}$$

is not always an exact expression. To obtain equation (17), for example, one has to put $\hat{U}(\vec{r},t)=1$ during all the time evolution of the ionized electron. This can be deduced from the first line of equation (12) (where we put $\hat{U}^+ = 1$) and equation (8) (where we put $H_I = H_I^{pA}$). However, as we have noted above, $\hat{U}(\vec{r},t)=1$ is not satisfied in strong laser fields. Burlon *et al* [10] and Leone *et al* [11] long ago considered in fact the same problem (nonresonant multiphoton ionization) in the $S$-matrix theory. With the help of some approximations done in the Green function representation of the $S$-matrix element they arrived at the same conclusion. They found that if one puts $\hat{U}(\vec{r},t)=1$ for all $t$ then one obtains equation (17). Burlon *et al* and Leone *et al* found that ionization rates based on equation (17) may depart by orders of magnitude from gauge-invariant ionization rates at the same level of accuracy in an analytical approximation to $\Psi_f^{(-)}$.

But there is another way of obtaining equation (17). If one assumes that the condition (1) is fulfilled then

$$\lim_{t \to \pm\infty} \hat{U} = 1, \quad \text{and} \quad \lim_{t \to \pm\infty} \hat{U}^+ = 1. \tag{18}$$

If so, one can drop $\hat{U}$ and $\hat{U}^+$ in all the equations (10), (11). Hence the ionization probability amplitudes (17) and (15) become equal. Of course, the respective direct time probability amplitudes of ionization become equal in this case too. The supposition, that for the accurate wavefunctions the LG and the VG KFR theories would give identical results, was called "strong gauge invariance" (see [4], p. 20). We have just shown that this is true, if the vector potential of the laser field vanishes at asymptotic times. The latter condition implies (equations (1), (2)) that

$$\int_{-\infty}^{\infty} \vec{F}(t)dt = 0 \ , \tag{19}$$

which is valid for a laser pulse [41] (see also [43], p. R165).

By the way, let us note that the well-known result for the atomic photoeffect (for more details see [44,45]) can be deduced as a conclusion from our considerations. For such process only one photon of energy $\hbar\omega > E_B$ ($E_B > 0$ - binding energy) is absorbed and the ionizing radiation is so weak, that it can be treated as a perturbation. Therefore one can substitute $\hat{U}(\vec{r},t) = 1$ for all $t$, obtaining from equation (8)

$$(S-1)_{fi} = \frac{1}{i\hbar}\int_{-\infty}^{\infty} dt \left(\Psi_f^{(-)dE}, H_I^{dE}\Phi_i\right) = \frac{1}{i\hbar}\int_{-\infty}^{\infty} dt \left(\Psi_f^{(-)pA}, H_I^{pA}\Phi_i\right) . \tag{20}$$

In the dipole approximation, in the limit of very low radiation intensities, in the exact wavefunctions $\Psi_f^{(-)dE} = \Psi_f^{(-)pA}$ one can put $\vec{A}(t) = \vec{0}$, so they contain no laser field. Thereby both of them become stationary Coulomb waves (of positive total energy, with an asymptotic momentum $\vec{p}$ as a parameter), which are orthogonal to $\Phi_i$. As a result, one obtains that the VG $S$-matrix element (the $\vec{A}^2$ term may be neglected) and the LG $S$-matrix element are identical, so the respective cross sections also do, if we use the accurate atomic wavefunctions.

Of course, from the practical point of view, the much more interesting question is what gauge we should choose, if the Volkov wavefunction is used for the final continuum atomic state, instead of the exact wavefunction. This problem does not have any general solution for today, but there are several examples (see section VI of [21]) showing that the approximate $S$-matrix theory in the version of Keldysh (LG) is in many cases better than the analogical approximate $S$-matrix theory in the version of Reiss (VG). The Keldysh theory is usually more accurate, but sometimes it may be even the only one, which is in qualitative agreement with the results of numerical solution to the TDSE [19]. In the low-frequency-high-intensity limit ionization rates calculated in the Keldysh theory for the $H(1s)$ atom are always closer to other theoretical results than their counterparts calculated in the Reiss theory (we refer the reader to our recent papers [21,23] for more details). In the recent experiment of Buerke and Meyerhofer [46] ionization rate of $He^+(1s)$ in the circularly polarized laser field of very low frequency $\omega \cong 0.043$ $a.u.$ ($\lambda \cong 1.053$ $\mu m$; we use atomic units here: $\hbar = e = m_e = 1$) has been measured with the relative error of roughly $35\%$. The only KFR result, which is in quantitative agreement with this measurement, is the WKB Coulomb corrected Keldysh ionization rate, which is about $15\%$ larger [23]. For the $H(1s)$ atom in an arbitrary static electric field (of strength $F$) ionization rate can be computed exactly by using the complex scaling method [47,48]. In the limit $\omega \to 0$, for the circularly polarized laser field, the above threshold ionization rates should approach a limit of exact static field ionization rates. One may also expect that in the limit $\omega \to 0$, for the linearly polarized laser field, the above threshold ionization rates would approach a limit of exact static field ionization rates averaged over one field period. For the Floquet theory the last two statements have been confirmed (with some provision regarding intermediate resonances for linear polarization) for not too strong fields [49]. In a way the fact that for the $H(1s)$ atom in the linearly polarized low-frequency laser field ionization rate can be treated as a cycle-averaged static-field ionization rate has been confirmed in *ab initio* simulations for $F \geq 0.1$ $a.u.$ (roughly) and frequencies $0.0456$ $a.u.$ $\leq \omega \leq$ $0.182$ $a.u.$ (see figure 2 and the discussion in [47]). We expect the same for the correct KFR theory, at least approximately (because in practice the KFR theory can be only approximate), in the limit of strong fields ($z_1 \to \infty$).

In figure 1 we compare the Keldysh and the Reiss ionization rates (as a function of the electric field strength $F$) for the circularly polarized laser field of frequency $\omega = 0.01$ $a.u.$ with the exact static field results of Scrinzi *et al* [47,48]. There is a very considerable difference, up to more than six orders of magnitude, between the Keldysh and the Reiss theories for the strongest fields shown in figure 1 ($F = 1$ $a.u.$ corresponds to $z_1 = 20000$, which is beyond the conventional applicability limit of the nonrelativistic theory, the latter being at $z_f = 2U_P/c^2 = 0.1$ [3], which corresponds to $F \cong 0.433$ $a.u.$). As expected, the gauge-invariant Keldysh theory, which utilizes the Volkov wavefunction (without any Coulomb corrections; equation (27) of [21] describes ionization rate in this case) improves with increasing intensity. In contrast, the Reiss theory gives unexpected decreasing values of ionization rate above $F \cong 0.2$ $a.u.$. For the highest intensities shown in figure 1 the Keldysh ionization rates are able to reproduce the exact static field ionization rates within the factor of 2. Ionization rates of Landau [50] (see also equation (1) of [21], which become accurate in the limit $F \to 0$) are also shown for comparison in figure 1. In figure 2 we show analogous ionization rates, but for linear polarization ($F = 1$ *a.u.* corresponds to $z_1 = 10000$, and $z_f = 0.1$ corresponds to $F \cong 0.613$ *a.u.*). Static field ionization rates have been averaged over one field period (equation (2) of [21] describes the averaged Landau result). To this end we have made the B-splines interpolation of the exact static field results of Scrinzi *et al* [47,48] and then we have numerically calculated the respective integral (these averaged exact static field ionization rates are shown analogically by full circles, as in figure 1). In figure 2 the Keldysh ionization rates (equation (20) of [1]) approach the averaged exact static field ionization rates with increasing intensity. In contrast, the Reiss ionization rates remain at least two orders of magnitude smaller than one could expect, even for very strong fields.

In our opinion, the qualitative difference (particularly distinct for circular polarization) between the Keldysh and the Reiss theories in the low-frequency-high-intensity limit, is connected with the matter of gauge invariance. When calculating the probability amplitude of ionization in the $S$-matrix theory, one integrates the time derivative of the overlap $\left(\Psi_f^{(-)}, \Phi_i\right)$, which is a function of time. As equation (8) clearly shows, this integration is performed from the point $t = +\infty$, where we know the overlap: $\left(\Psi_f^{(-)}, \Phi_i\right) = \delta_{fi}$, to the point $t = -\infty$, where this overlap is (analytically) unknown. To obtain an approximate result, we approach the value of the integrand for all times between $t = +\infty$ and $t = -\infty$ by using the Volkov wavefunction instead of $\Psi_f^{(-)}$. But for any time $t$ obeying $+\infty > t > -\infty$ and $\vec{A}(t) \neq \vec{0}$ the overlap $\left(\Psi_f^{(-)dE}, \Phi_i\right)$ is the instantaneous probability amplitude of ionization and the overlap $\left(\Psi_f^{(-)pA}, \Phi_i\right)$ is not. These two expressions are equal only when $\vec{A}(t) = \vec{0}$. Therefore only in the LG our approximation has a clear physical interpretation for all $t$, which contribute to the analytical value of the integral.

In conclusion, on the faith of a widely-accepted assumption, by a straightforward analytical calculation based on the TDSE, we have shown that the LG form of the exact $S$-matrix element (describing ionization in a strong laser field) is in fact the gauge-invariant form. This assumption says that having the solution of the TDSE in the VG, one has to multiply it by the factor of $\exp\left(\left(-ie/\hbar c\right)\vec{r}\vec{A}(t)\right)$ before projecting it on an eigenstate of the atomic Hamiltonian (and no additional factor is needed in the LG). In this paper traditionally we were using the name of "the VG form of the exact $S$-matrix element" for equation (17).

We have shown, that in fact one should rather use this name for equation (15), where $\vec{F}(t)$ is expressed through $\vec{A}(t)$. The $S$-matrix element from equation (17) can be equal to its counterpart from equation (15), if one of the two following conditions is fulfilled: (i) the Göppert-Mayer exponential factor is approximated by unity during all the time evolution of the ionized electron or (ii) the vector potential of the laser field vanishes at asymptotic times.

**Acknowledgments**

The author is extremely grateful to Professor Piotr Kosiński for numerous interesting and enlightening discussions and remarks connected with this paper, and for reading the manuscript. The author is also grateful to Professors Howard R. Reiss, Andrzej Raczyński and Jarosław Zaremba for their interesting correspondence on the subjects related to this paper. The present paper has been supported by the University of Łódź.

**Appendix**

Unlike the steps leading from equations (10), (11) to equations (15), (16) (which to the best of our knowledge have not been explicitly shown so far) this appendix is partly based on considerations similar to those present in the reach literature of the subject, for example [31-33,36-38,51]. Here we put forward an independent argument for the form of the asymptotic reference states, which appear in the $S$-matrix element. From self-evident reasons some quantities in this appendix have a different notation.

If a total duration time of a laser pulse is $2\tau$, the following electromagnetic field interacts with an atom in the dipole approximation: the magnetic field $\vec{B}(\vec{r},t) \equiv \vec{0}$ (for all $\vec{r},t$) and the electric field $\vec{E}(\vec{r},t) \equiv \vec{F}(t)$ (for all $\vec{r},t$, where $\vec{F}(t)$ is a certain real continuous function vanishing outside the time interval $[-\tau,\tau]$). Since $\vec{B} = \vec{\nabla} \times \vec{A}$ and $\vec{E} = -\vec{\nabla}\phi - \frac{1}{c}\frac{\partial \vec{A}}{\partial t}$, the most general vector and scalar potentials describing these fields are

$$\vec{A}(\vec{r},t) = \vec{\nabla}\chi(\vec{r},t), \quad \text{and} \quad \phi(\vec{r},t) = -\vec{F}(t)\vec{r} - \frac{1}{c}\frac{\partial}{\partial t}\chi(\vec{r},t), \qquad (A1)$$

where $\chi(\vec{r},t)$ is an arbitrary real differentiable gauge function. Of course, physical predictions formulated in any particular gauge $\chi$ should not depend on $\chi$. Utilizing the well-known usual procedure of the minimal electromagnetic coupling, we write

$$H_\chi = \frac{1}{2m}\left(\vec{p} - \frac{e}{c}\vec{A}\right)^2 + eV + e\phi, \quad \text{and} \quad i\hbar\frac{\partial}{\partial t}\Psi_\chi(\vec{r},t) = H_\chi \Psi_\chi(\vec{r},t). \quad (A2)$$

The gauge-dependent total Hamiltonian of the atom in the laser field $H_\chi$ determines the time evolution of the electronic wavefunction in the TDSE. In equations (A2) $V = V(\vec{r})$ is the binding potential (for example the Coulomb one), and $\vec{A}, \phi$ are given by equations (A1). The well-known relationship exists between Hamiltonians and wavefunctions in different gauges [25-27]. In particular, if one of the gauge functions is zero (we put the index "0" for $\chi(\vec{r},t) \equiv 0$) we have

$$H_\chi = \exp(ie\chi(\vec{r},t)/\hbar c) H_0 \exp(-ie\chi(\vec{r},t)/\hbar c) - \frac{e}{c}\frac{\partial}{\partial t}\chi(\vec{r},t),$$

(A3)

$$\text{and} \quad \Psi_\chi(\vec{r},t) = \Psi_0(\vec{r},t)\exp(ie\chi(\vec{r},t)/\hbar c).$$

If equations (A3) are satisfied, all the TDSE equations (A2), for different gauge functions $\chi(\vec{r},t)$, are equivalent.

Now let us consider the limits $t \to \pm\infty$, where $\vec{F}(t)$ is zero, but $\chi(\vec{r},t)$ - not necessary! When the laser field is off, the Hamiltonian $H_\chi$ is equal to $H_\chi^{atom}$. The latter one is given by equations (A1), (A2) and the condition $\vec{F}(t) \equiv \vec{0}$. $H_\chi^{atom}$ is gauge-dependent and its eigenvalues are also gauge-dependent, so this Hamiltonian should not define initial or final reference states. This is the term $\frac{e}{c}\frac{\partial}{\partial t}\chi(\vec{r},t)$, which breaks gauge invariance in the first equation (A3). Instead, the gauge-invariant (in a sense described below) energy operator $\varepsilon$ should be used to define these states. The energy operator $\varepsilon$ is a sum of the kinetic and potential energy operators and $\varepsilon_0 = H_0^{atom} = \vec{p}^2/2m + eV$ in the gauge defined by $\chi(\vec{r},t) \equiv 0$. In the arbitrary gauge one can obtain this operator from $H_0^{atom}$ by the relation

$$\varepsilon_\chi = \exp(ie\chi(\vec{r},t)/\hbar c) H_0^{atom} \exp(-ie\chi(\vec{r},t)/\hbar c) = \frac{1}{2m}\left(\vec{p} - \frac{e}{c}\vec{\nabla}\chi(\vec{r},t)\right)^2 + eV \ . \tag{A4}$$

Suppose that in the following eigenequation: $\varepsilon_0 \Phi_0(\vec{r},t) = E\Phi_0(\vec{r},t)$ $E$ (the total energy) is a certain real eigenvalue of $\varepsilon_0$. Transforming this eigenequation (using equation (A4) and the equation: $\Phi_\chi(\vec{r},t) = \Phi_0(\vec{r},t)\exp(ie\chi(\vec{r},t)/\hbar c)$) to the arbitrary gauge $\chi(\vec{r},t)$ one obtains

$$\varepsilon_\chi \Phi_\chi(\vec{r},t) = E\Phi_\chi(\vec{r},t) \ . \tag{A5}$$

Therefore the eigenvalue $E$ remains the same in all gauges, and this is why we call the energy operator $\varepsilon_\chi$ (or $\varepsilon$ in brief) gauge-invariant. Although it depends on the gauge function $\chi(\vec{r},t)$, equation (A5) holds true for any $\chi(\vec{r},t)$. To define both equivalent forms (the direct time and the reversed time) of the exact $S$-matrix, we need the initial and final $(i,f)$ asymptotic reference states (see equations (6), (7)), which should obey equations of the type (A5) in the arbitrary gauge. In the photoionization or photodetachment process bound-free transitions occur, so $E_i < 0$ and $E_f > 0$. From the above discussion it follows clearly that choosing the initial and final states as eigenstates of $H_0^{atom} = \vec{p}^2/2m + eV$ corresponds to the gauge defined by $\chi(\vec{r},t) \equiv 0$. In other gauges the initial and final states take the form:

$$\Phi_\chi^{initial}(\vec{r},t) = \Phi_0^{initial}(\vec{r},t)\exp(ie\chi(\vec{r},t)/\hbar c) \ ,$$

$$\Phi_\chi^{final}(\vec{r},t) = \Phi_0^{final}(\vec{r},t)\exp(ie\chi(\vec{r},t)/\hbar c) \ , \tag{A6}$$

where $\Phi_0^{initial}(\vec{r},t)$ and $\Phi_0^{final}(\vec{r},t)$ are the well-known "textbook" wavefunctions (for example the Coulomb ones). They are products of some functions of $\vec{r}$ only and the exponential factor $\exp(-iE_i t/\hbar)$ or $\exp(-iE_f t/\hbar)$ respectively. In the definitions of the $S$-matrix elements in equations (6) and (7) there are two different limits $t \to -\infty$ and $t \to +\infty$. Therefore it follows from equations (A6) that in the arbitrary gauge $\chi(\vec{r},t) \neq 0$ one may not simply drop the time evolution (from $-\infty$ to $+\infty$ or reversely) of the factor $\exp(ie\chi(\vec{r},t)/\hbar c)$.

Finally, it follows from equations (A1) and (A2) that the gauge defined by $\chi(\vec{r},t) \equiv 0$ is the LG, because the total Hamiltonian of the atom in the laser field takes the form:

$H_0 = \vec{p}^{\,2}/2m + eV(\vec{r}) - e\vec{F}(t)\vec{r}$. On the other hand, one can choose the gauge function $\chi(\vec{r},t)$ such that $-\frac{1}{c}\frac{\partial}{\partial t}\chi(\vec{r},t) = \vec{F}(t)\vec{r}$ (this condition defines some infinite set of gauge functions $\chi(\vec{r},t) \not\equiv 0$), obtaining

$$\chi(\vec{r},t) = -c\vec{r}\int_{-\infty}^{t}\vec{F}(t')dt' + g(\vec{r}) , \tag{A7}$$

where $g(\vec{r})$ is an arbitrary function. Substituting this into equations (A6) and utilizing the definition of $\vec{F}(t)$ one obtains

$$\Phi_\chi^{initial}(\vec{r},t) = \Phi_0^{initial}(\vec{r},t)\exp(ieg(\vec{r})/\hbar c) \quad \text{for } t < -\tau ;$$

$$\Phi_\chi^{final}(\vec{r},t) = \Phi_0^{final}(\vec{r},t)\exp\left(-(ie/\hbar)\vec{r}\int_{-\tau}^{\tau}\vec{F}(t')dt' + ieg(\vec{r})/\hbar c\right) \quad \text{for } t > \tau . \tag{A8}$$

From equations (A8) it is obvious that in the gauge $\chi(\vec{r},t) \not\equiv 0$ the initial and final reference states usually acquire different exponential factors (multiplying the "textbook" wavefunctions) with phases, which do not vanish even at $t \to \pm\infty$. Moreover, for an arbitrary function $\vec{F}(t)$, phases at $+\infty$ and $-\infty$ are different, unless

$$\int_{-\tau}^{\tau}\vec{F}(t)dt = \vec{0} . \tag{A9}$$

It follows from equations (A1) and (A2) that the gauge given by $\chi(\vec{r},t) = \vec{r}\vec{A}(t)$ (where $\vec{A}(t)$ is defined by the function $\vec{F}(t)$ and equation (2)) is the VG, because the total Hamiltonian of the atom in the laser field takes the form $H_\chi = (\vec{p} - e\vec{A}(t)/c)^2/2m + eV(\vec{r})$. Switching off the vector potential at asymptotic times (equation (1)), that one has to do in the $S$-matrix formalism, is possible only when equation (A9) is satisfied.

There is a proof (which does not utilize the dipole approximation) in appendix A of [52] showing that for any finite laser pulse the condition (A9) is satisfied everywhere inside the laser cavity. This constraint on the electric field of a laser pulse has been derived from the Maxwell equations provided that $\vec{\nabla}\vec{P} = 0$ ($\vec{P}$ denotes the polarization of the laser medium), what is true to a good approximation [53]. Our paper shows that Eq. (A9) is also the necessary condition on which the (usually) so-called exact VG probability amplitude is equal to the (usually) so-called exact LG probability amplitude (equation (20)). But the latter one is in fact the gauge-invariant result, which does not need any additional conditions, like (A9), to be valid. Perhaps in the real experimental conditions of strong laser field the condition (A9) could be satisfied only approximately ($0 < \left|\int_{-\tau}^{\tau} \vec{F}(t)dt\right| << 2\tau \max|\vec{F}(t)|$) and the additional phase which appears in the second of equations (A8) could not be neglected. Then the initial and final reference states (see equations (A8)) would have different phases in any gauge $\chi(\vec{r},t) \not\equiv 0$. This would be the source of a certain additional error not connected with any analytical approximations in the $S$-matrix theory. Only the gauge-invariant $S$-matrix element (equations (15) and (16)), which is equivalent to the one with $\chi(\vec{r},t) \equiv 0$, does not have this drawback.

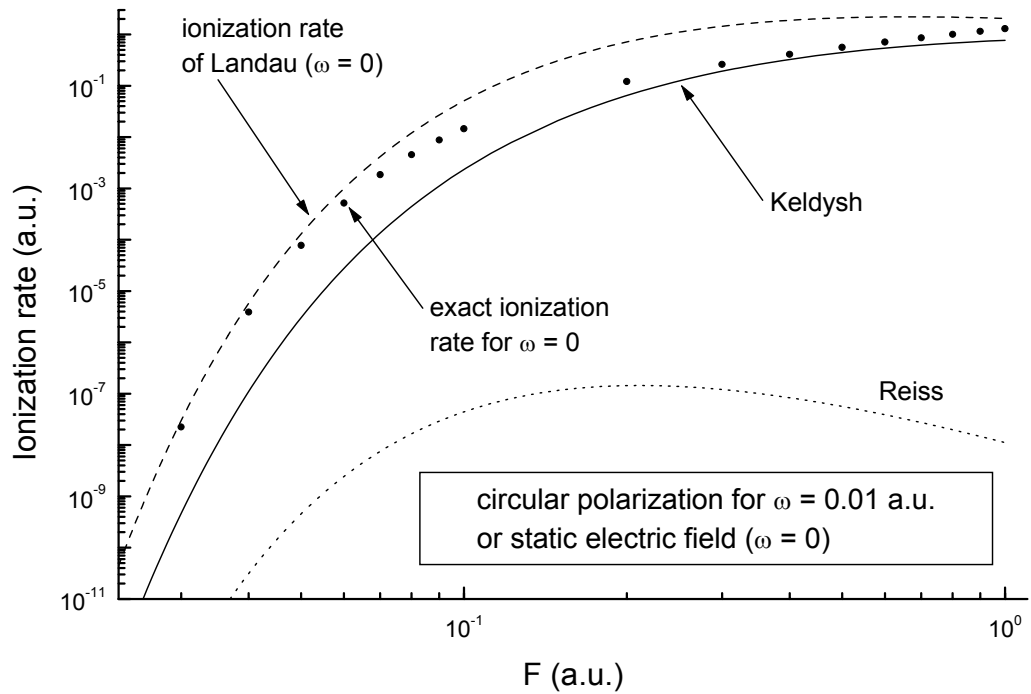

**Figure 1.** Ionization rates of the $H(1s)$ atom in the circularly polarized laser field (for $\omega = 0.01$ *a.u.*) or in the static electric field against the electric field (see text for details).

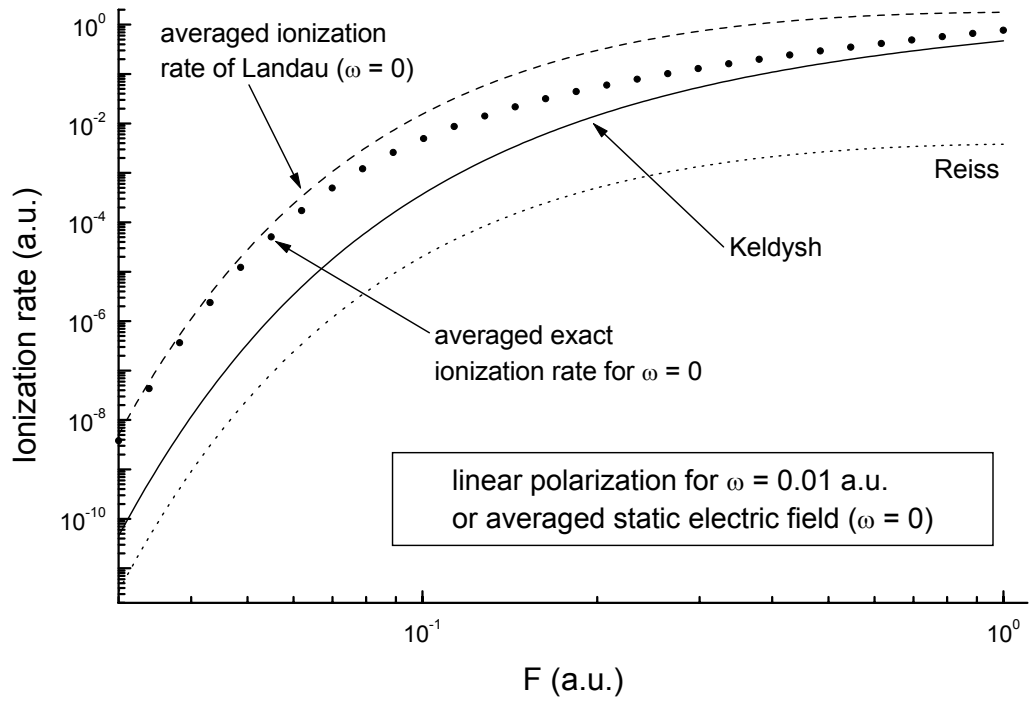

**Figure 2.** Ionization rates of the $H(1s)$ atom in the linearly polarized laser field (for $\omega = 0.01$ *a.u.*) or cycle-averaged static field ionization rates against the electric field (see text for details).